# LabOSat as a versatile payload for small satellites: first 100 days in LEO orbit

G. A. Sanca (1), M. Barella (2,3,4), F. Gomez Marlasca (4), G. Rodríguez (2), D. Martelliti (2), L. Patrone (2), P. Levy (3,4), F. Golmar (1,2,3)

(1) ECyT-UNSAM, Martín de Irigoyen 3100, B1650JKA, San Martín, Bs As, Argentina
(2) CMNB-INTI, Avda. Gral. Paz 5445, B1650KNA, San Martín, Bs. As., Argentina
(3) Consejo Nacional de Investigaciones Científicas y Técnicas (CONICET), Bs. As., Argentina
(4) DMC, GAIANN, CAC-CNEA, Avda. Gral. Paz 1499, 1650, San Martín, Argentina
*gsanca@unsam.edu.ar

Abstract: In this work the first results obtained by LabOSat-01 platform are presented. This platform was designed for testing custom devices on board of small satellites. Two LabOSat-01 type boards were launched and placed into Low Earth Orbit (LEO) on May 30, 2016. We present here an analysis of data collected by one of these boards during the first days of mission. Total Ionization Dose results are compared with data acquired by LabOSat-01's predecessor board, MeMOSat-01, launched in 2014.

**1. Introduction**

LabOSat (acronym for "Laboratory On a Satellite") is a universal electronic platform designed to perform experiments in harsh environments, such as outer space [1]. LabOSat-01, the first version of the platform, was designed for testing custom devices on board of small satellites. Two LabOSat-01 type boards were launched and placed into Low Earth Orbit (LEO) on May 30, 2016 in two different nano-satellites, Ñusat-1 and Ñusat-2, developed by the Argentinian company Satellogic [2].

The LabOSat-01 board is divided into different subsystems. The "MeMO" subsystem is dedicated to experiments on custom made ReRAM devices, a novel technology based on the Resistive Switching phenomena on metal-oxide-metal capacitor like devices. Another subsystem, called "xFET", was specially designed to test any transistor device. The goal of these two subsystems is to test devices' degradation under hostile conditions. Stimuli applied to Devices Under Test (DUTs) can be either sourcing voltage or current. The MeMO subsystem allows operation over 30 devices while the xFET subsystem allows over 7. Both of them work as Source-Measure units (SMUs).

Although the satellites have been in orbit since June 2016, first experiments began in November after satellite's boot-up routines. After more than 100 days of operation in LEO, telemetry data and results from the experiments running inside LabOSat-01 are analyzed and discussed.

**2. Background: MeMOSat-01**

LabOSat's predecessor, MeMOSat-01, is a CubeSat-like platform which was specially designed to measure ReRAM devices in LEO. It was installed inside BugSat-1 and is working since June 2014. Besides measuring the I-V curves of the DUTs, MeMOSat-01 senses Total Ionization Dose (TID) using COTS pMOS transistors. The principle of operation and the design of those subsystems in LabOSat were inherited from MeMOSat-01, nonetheless they were improved to cope with new requirements.

**3. Brief description of the platform**

**3.1 Hardware**

LabOSat-01's communication and experiment modules are commanded by the ultra-low-power microcontroller (MCU) MSP430F1612, frequently used in LEO CubeSat-like missions. Guertin *et al*. [3] showed that this MCU presented no failures up to 20 krad of Total Ionization Dose. Calculations performed by Samwel *et al*. [4], [5] suggest that expected TID for LEO must be at maximum 10 krad/yr and strongly decreasing depending on the Al shielding. At the present time, measurements realized by MeMOSat-01 in BugSat-1 [also presented in this proceeding book – Lipovetzy et al.] show that TID is approximately 1.2 krad/yr, at a 480 rad/yr rate, which implies that the MCU is suitable for LabOSat's 3-year mission.

LabOSat's SMUs are capable of exciting the DUTs with voltage or current. MCU's 12-bit DACs outputs are used to control voltage or current sources that force on-board devices. The 12-bit ADCs inputs are used to measure the response of the DUTs and other parameters that are periodically checked to guarantee the correct behaviour of the board. The DACs and ADCs are configured to use the internal 2.5 V voltage reference. To increase the dynamic range in measurements and avoid out-of-range ADC readings, amplifier or attenuator stages are used. GPIO ports are used to switch between subsystems or modes of operation.

The MeMO subsystem was designed to perform experiments over memory devices, particularly ReRAM devices, a technology believed to be intrinsically resistant to radiation [6]. In general, any 2-terminal device can be excited and measured. Another experiment is running aboard LabOSat, designed to study the performance of transistors. This subsystem is named xFET and basically is able to test any 3-terminal device, commercial or custom made.

A more detailed description of the LabOSat's subsystems, like ReRAM and xFET modules, and condition-signal measurement can be found in a previous work made by Barella *et al.* [1].

**3.1.2 Dosimeters**

LabOSat-01 has COTS pMOS transistors, characterized before integration to measure TID. Its exposure to ionizing radiation causes, among other effects, a shift of the threshold voltage (Vth). This effect is explained by charge accumulation in the traps of an oxide. Electron-hole pairs are created when a charged particle passes through the gate oxide of MOS transistors. In silicon oxide, electrons has a higher mobility than holes so they are swept out of the oxide, typically in a picosecond or less. However, in that first picosecond, a fraction of the electrons and holes will recombine. The holes, which escape initial recombination, are relatively immobile and remain near their point of generation, where they cause a negative threshold voltage shift in a MOS transistor [9].

Hence, as Vth is affected by TID, its shift can be used to estimate the dose absorbed by the board. This parameter is measured indirectly by biasing the source of the pMOS transistor with both known current and voltage (IS and VS) while the drain is grounded. Then, reading the gate voltage allows to observe variations in VGS, which will be traduced to Vth variations [7], [8].

Figure 1 shows a simplified schematic of the dosimeter's configuration. LabOSat has four on-board dosimeters and the possibility to connect other four external dosimeters, to be located in strategic spots. In order to measure all of them, multiplexers are used. In Ñusat missions, only two on-board dosimeters were placed in each board. Also, a resistor was welded as control.

As shown in the schematic, the voltage is measured at the gate of the transistor. Since it is a pMOS transistor, drain is grounded and VGS voltage which is used to measure TID must be negative. In this configuration, the source voltage value is determined by the voltage divider formed by R1 and R2 resistors. Thus, the reference voltage, VREF, is a fraction of the supply voltage.

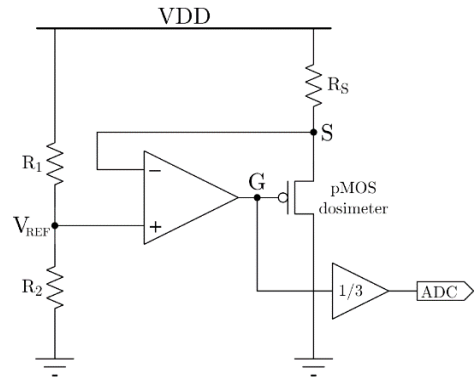

**Figura 1**. Simplified schematic of the dosimetry module.

As the gate voltage must exceed 2.5V, it is out of the ADC reading range. To solve this, an attenuator stage is used, as shown in the Figure 2. Finally, VGS is determined by

$$V_{GS} = V_G - V_S = 3 V_{ADC} - \alpha V_{DD} \quad (1)$$

here α is a fraction given by the voltage divider. The variation in time of the voltage VGS will be proportional to the incident TID, according to the sensitivity of each dosimeter. To improve accuracy in the measurement, corrections using measured temperature and power supply are possible.

**3.2 Firmware**

Firmware was designed to perform three major tasks: communicate with the satellite's main computer, control the experiments and process data acquired during the experiments. Communication with the satellite, physically implemented with a 200 kHz 3-wire SPI interface, allows LabOSat-01 to receive command lists and to send data packages. LabOSat-01 is either waiting for a new command list (as an idle state) or executing one. There are two types of command lists: Arbitrary Lists (generated on the ground, and then uploaded) and Standard Test Requests (automatically generated by the satellite).

Arbitrary Lists are meant to command custom experiments and/or firmware maintenance chores. The firmware embedded in LabOSat-01 is capable of processing these Arbitrary Lists, which can be put together to perform a wide variety of tasks; e.g., updating/fixing firmware, perform different experiments on the DUTs, ask for telemetry data, run self-integrity tests, etc.

Standard Tests are run, at most, one time a day. The satellite is programmed to send Standard Test Requests periodically to LabOSat-01; if one day has passed since the last Standard Test execution, LabOSat-01 allows the request and executes a new Standard Test. Standard Tests consist of a set of preconfigured experiments and activities that must be sequentially executed.

Among the scheduled activities there are battery voltage measurements, temperature and dosimetry readings, firmware self-integrity checks (using a 16-bit CRC routine), etc. Firmware in LabOSat-01 was tailored to execute different kinds of standard experiments in every type of DUT onboard. The single-bit ReRAM memory banks are characterized by performing I-V curves measurements and endurance tests. xFET transistors are characterized with different families of I-V curves.

## 4. Devices Under Test

### 4.1 Single-bit ReRAMs

Tested custom ReRAM devices are tri-layer structures composed of two metallic electrodes divided by a thin film oxide fabricated at our laboratories. The studies are focused on non-stoichiometric titanium oxide ($TiO_{2-x}$) and manganite $La_{1/2}Ca_{2/3}MnO_3$. The devices' functionality rely on the Resistive Switching phenomenon. When an electric field is applied cross the dielectric it produces mixed filamentary structures formed by vacancies of the oxide cell and metal atoms from the electrodes. This structure reduces the resistance of the dielectric oxide cell, modeled as a dielectric plus a parallel resistance of lower value. In some cases, depending on the applied electric field, the filamentary structure could short circuit the electrodes and leave the device in a very low resistance state.

The switching between different non-volatile resistance values or "states" is exhibited when this filamentary structure is disarmed (or even fused) and formed again. This is accomplished by applying voltage or current pulses of opposite polarity (bipolar Resistive Switching).

### 4.2 Thin Film Transistors

Into LabOSat-01's xFET module there are ZnO based Thin Film Transistors (TFTs) developed by E. Lopez *et al*. [E. Lopez 2016]. TFTs are three terminal field effect devices with a structure assembled by deposition of different types of thin films such as gate, gate insulator, semiconductor channel and source-drain contacts onto a substrate.

TFTs operation could be compared with classic MOSFET operation. The threshold voltage concept could be applied to TFTs too and operation regions are the same: cut-off, linear or triode and saturation. The analysis of the current-voltage characteristics of TFTs are similar to those for conventional MOSFETs. The goal is the in-orbit observation of $V_{th}$ shifts in correlation with TID. The main hypothesis is that the film gate oxide (25 nm) will not be able to collect charge so a change in the threshold voltage is unlikely to occur.

To determinate the threshold voltage, the saturation region is taken into account. The drain current, $I_D$, depends quadratically on the gate-source voltage, $V_{GS}$. By linearizing, the relationship between $I_D$ and $V_{GS}$:

$$I_D = k(V_{GS} - V_{th})^2 \quad (2)$$

$$\sqrt{I_D} = \sqrt{k}\, V_{GS} - \sqrt{k}\, V_{th} \quad (3)$$

where $\sqrt{k}$ is the slope and $-\sqrt{k}\, V_{GS}$ the intercept.

## 5. Results

Presented results in this work are partial results. The experiments are intentioned to last the 3-years mission time. The elapsed time is not significant in order to conclude nothing about DUTs performance in LEO. Although, it is important to remark this preliminary results are useful to improve our development process, both with LabOSat systems and with ReRAM devices.

### 5.1 Report's chronology

In the present work, collected data between February 3, 2016 and February 28, 2017 is analyzed. The first four reports (listed from 3 to 6) were generated in our labs, in order to test the correct platform functioning before delivering the board for integration into satellite. Reports number 7, 8 and 9 were generated after integration and before launching. In Figure 2, report's chronology is represented. In brackets report numbers are presented corresponding to each milestone, along with a brief description.

The lift-off took place on May 30, 2016. After four months, on September 7, LabOSat was switched on for the first time in LEO at 498km (average) and tests started to run periodically. Data generated by the platform is downloaded within satellite's telemetry. This information include temperature, GPS coordinates, power supply, Cyclic Redundancy Check (CRC), dosimetry, response of the DUTs.

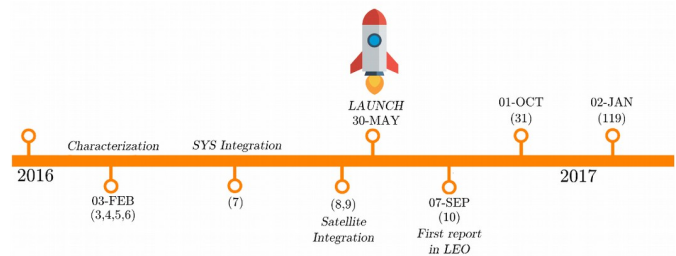

**Figura 2**.LabOSat's timeline 2016-2017. Milestones (report number).

### 5.2 Operating conditions

#### 5.2.1 Temperature

Temperature measurements are performed both by LabOSat platform and the satellite. The sensor reports temperature with a resolution of 1ºC in range between -50ºC and 127ºC. Future missions will improve the resolution. Temperature measurements taken before the end of a Standard Test routine is always higher than the temperature at the beginning of it.

Satellite's temperature measurements showed a mean increase of $(5.03 \pm 0.14)$ºC between values obtained before and after the Standard Test routine. This increase in temperature is explained by the board warming-up due to power consumption and dissipation.

#### 5.2.1 Power Supply

In order to perform corrections of $V_{REF}$ variations in dosimetry module, ReRAM and xFET subsystems, it is essential to know exactly the supply voltage of the platform. The battery voltage values are plotted against report number in Figure 4. The circled red bullets represent the measurement at the start of the standard test, while the squared blue bullets are the measurements in the middle of the routine.

Large variations between days/reports are due to power cycles of the satellite. Within a test, supply voltage presents small drops which make clear that monitoring this key variable is crucial to perform the mentioned corrections.

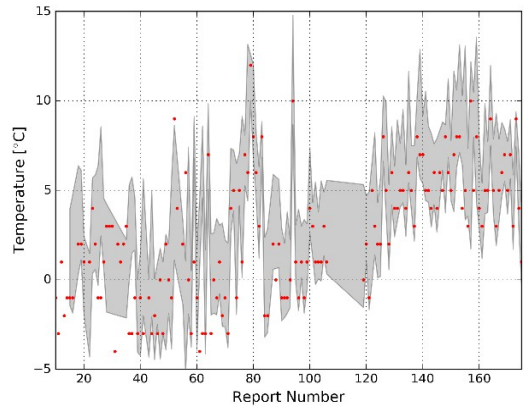

**Figura 3**. Temperature measurements performed by LabOSat and satellite. Upper bound of the gray area represents temperature after a Standard Test and the lower bound represents the temperatures before the test starts. Red dots: temperature measured by LabOSat.

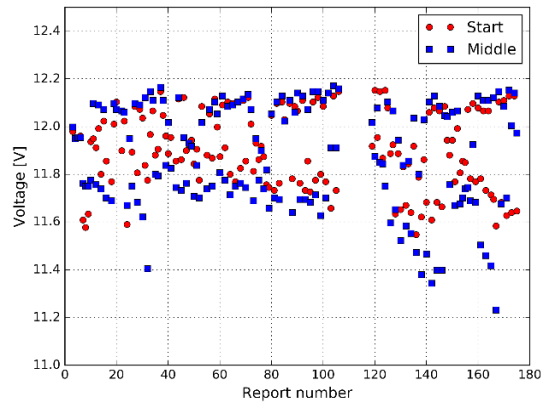

**Figura 4**. Supply voltage measurements performed by LabOSat.

### 5.2.1 Dosimetry

According to the simulations of Samwel *et al*., TID for LEO with a 1.5 mm thick of Al shielding, which is similar to the environment where LabOSat is working, will be 1 krad/yr at maximum. Assuming a constant dose during a year and considering the sensitivity of each dosimeter is expected that one dosimeter shifts $V_{GS}$ around 48 mV and the other one around 40 mV, for a period of 175 days. These values represent 79 and 66 ADC counts, respectively.

As showed in Equation 1, gate-source voltage measurement is performed indirectly. LabOSat senses the voltage at the gate relative to ground and then subtracts the value of a reference voltage which polarize the source pin. That reference is not independent of $V_{DD}$. In this way, the variations in the gate source voltage caused by $V_{th}$ shifting for the first days of the mission are masked by the variations of the supply voltage. Furthermore, no clear trend was observed in $V_{GS}$. It is important to remark that although the pMOS transistor is polarized in the Zero Temperature Coefficient (ZTC) point, VGS measurements were found to be uncorrelated with the temperature (results not showed here).

### 5.3 Memory devices

In this mission all ReRAM devices were configured to perform current-voltage (I-V) experiments so as to observe if resistive switching parameters were modified by environmental conditions. In first place, no degradation was observed over the current and voltage source. Control devices i.e. fixed resistance loads, allowed us to check that characteristics of sources remain the same as measured at our lab. Secondly, some devices were found to be in high impedance after several I-V loops.

After studying the pinout of damaged devices and the packaging properties like epoxy, wire-bonding and pads material, we concluded that high variations of temperature (T) might cause this failure. In spite of being inside the satellite LabOSat recorded temperature variations (ΔT) of 15ºC which could mechanically stress the bonded structure caused by dilation-contraction cycles. Nonetheless, other devices kept switching as shown in Figure 6. Different runs are plotted for titanium dioxide and manganite ReRAM. The $TiO_{2-x}$ devices presented hysteresis the first 30 days in orbit. After day 31, the issues related with encapsulation of the silicon die produced a permanent damage, which led to high impedance DUTs. The manganite devices continue to exhibit hysteresis and are currently switching between two states. However, switching parameters changed from our lab tests as can be seen in the I-V curves. We attribute this behavior to the intrinsic variations that suffer low endurance ReRAM devices. This effect may not be related to incident radiation because dosimetry does not measure significant TID.

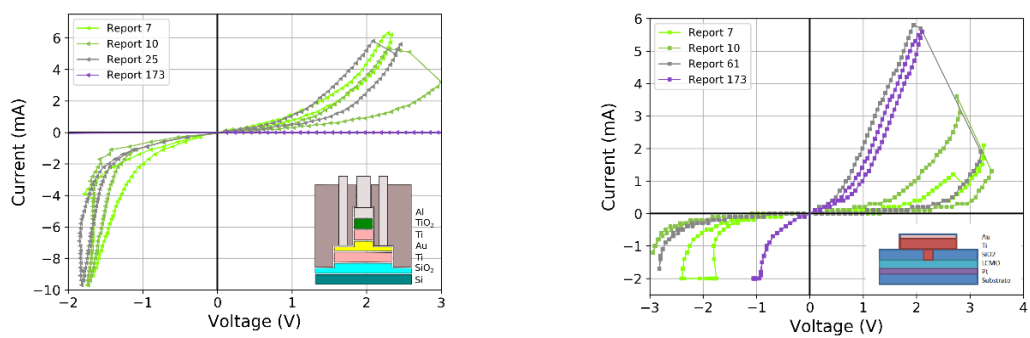

**Figure 6**. Current-voltage response of ReRAM devices for diferent days in LEO. Left: titanium dioxide ReRAM. Right: manganite ReRAM.

### 5.3 TFT devices

The analysis results of the first 175 days indicates that there is no observable $V_{th}$ shifting in the TFT devices. However, taking into account the dosimeters measurements, it is expected that the incident TID does not generate changes in the threshold voltage yet. This is just a partial analysis since it is a 3 years mission. Nevertheless, we expected that the thin gate oxide will not accumulate electrical charge so $V_{th}$ shifting will not occur.

In Figure 7, the $I_D$ vs. $V_{DS}$ curve for one device is shown for different reports. Also, threshold voltage was calculated from the characteristic output curve using Equations 2 and 3. The time evolution of this parameter does not exhibit a clear pattern, if there was one. Figure 7 shows the histogram of the extracted $V_{th}$ for the same device with an inset of its value as a function of report number.

### 6. Conclusions

LabOSat has been working in LEO for more than 170 days. Regarding performance, all subsystems are working as expected. On Earth calibration, supply voltage and temperature measurements in orbit allows us to carry out corrections to improve accuracy in DUTs experiments. TID measurements are still inconclusive, but consistent with the expected results, signaling very lows radiation levels. ReRAM devices still exhibit hysteresis. Variations of I-V curves are due to the low endurance instead of absorbed dose. It is important to remark that this is a 3-years mission, so the results presented here are partial results.

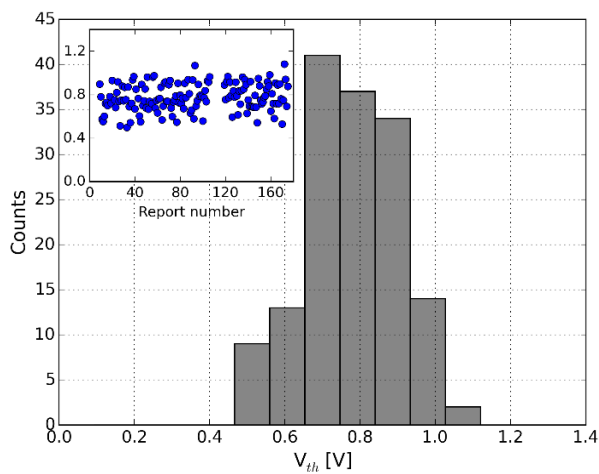

**Figure 7**. Histogram of the extracted threshold voltage for one device. Inset: time (report number) evolution of this parameter.

## 6. Acknowledgment

The authors would like to thank to R. Ferreira, E. Paz, W. R. Acevedo and D. Rubí who participated in the fabrication of ReRAM devices; P. Stoliar, L. Hueso and E. Lopez who were involved in fabrication of xFET devices; M. G. Inza who facilitated the dosimeters. The authors acknowledge financial support from ANPCyT PICT 2013-0788 "MeMOSat" and UNSAM-ECyT FP-001.